\documentclass[trackchanges]{aastex7}
\usepackage{amsmath}
\usepackage{comment}

\newcommand{\expAxi}[0]{x}

\begin{document}

\title{The metallicity of giant exoplanets around low-mass stars and how Ariel can help}

\author[0000-0001-5555-2652]{Ravit Helled}
\affil{Department of Astrophysics, University of Zurich, Winterthurerstrasse 190, CH-8057 Zurich, Switzerland}
\email[show]{ravit.helled@uzh.ch}

\author[orcid=0000-0001-5555-2652,sname='Sho Shibata']{Sho Shibata}
\altaffiliation{please}
\affiliation{Rice University}
\email{ravit.helled@uzh.ch}  

\author[0000-0001-8401-4300]{Shubham Kanodia}
\affil{Earth and Planets Laboratory, Carnegie Institution for Science, 5241 Broad Branch Road, NW, Washington, DC 20015, USA}\email{skanodia@carnegiescience.edu}

\author[0000-0002-5494-3237]{Billy Edwards}
\affil{SRON, Space Research Organisation Netherlands, Niels Bohrweg 4, NL-2333 CA, Leiden, The Netherlands}\email{b.edwards@sron.nl}


\begin{abstract}
Giant planets orbiting low-mass stars represent a unique population of giant planets that can be studied to constrain planet formation theory. Surveys of transiting giant exoplanets around M-dwarfs (GEMS) allow for measurement of their masses and radii, which in turn can be used to estimate their bulk densities. Coupling these observations to interior structure and evolution models can yield the bulk metallicity of these planets, however there are degeneracies to this that are improved by measurements of atmospheric metallicity. Estimates of bulk metallicity can be crucial to our understanding of planet formation timescales and mass budgets, particularly in these extreme (planet-to-star) mass ratio systems.

 Here we show that GEMS are expected to have a low bulk metallicity as a natural result of the planet formation process, and the low efficiency of planetesimal capture post-formation. To test this empirically, we need a larger sample of GEMS with atmospheric measurements, which can reduce some of the degeneracies with the interior modelling of planets. Measuring the atmospheric composition of GEMS with JWST and Ariel will be crucial for better understanding this unique planetary type, which sits at the tail of standard planet formation conditions.

 
\end{abstract}



\section{Introduction} 
Small stars, particularly M-dwarfs, constitute the most abundant stellar population in the universe, accounting for approximately 75\% of all stars in the solar neighborhood \citep{henry_solar_1994, reyle_10_2021}. Their prevalence, coupled with their extended main-sequence lifetimes—which can span trillions of years compared to the $\sim$10 billion-year lifespan of Sun-like stars—has positioned M-dwarf systems as crucial targets in the search for and characterization of exoplanets. Recent results from the Transiting Exoplanet Survey Satellite (TESS) and the James Webb Space Telescope (JWST) have significantly refined our understanding of the terrestrial population around these low-mass hosts \citep{ment_occurrence_2023}.

Demographic surveys indicate that terrestrial planets are exceptionally prolific around mid-to-late M-dwarfs \citep{dressing_occurrence_2015, hardegree-ullman_kepler_2019}. Furthermore, JWST observations of targets like the TRAPPIST-1 system and LHS 1140b can determine whether these planets can retain secondary atmospheres despite the intense XUV radiation of their hosts' youth \citep{Greene2023, JWST2026}.

However, in addition to these rocky planets, giant exoplanets around M-dwarfs (GEMS) present an interesting population with extreme planet-to-stellar mass ratios. Starting with Doppler measurements of non-transiting GEMS to the currently growing sample of transiting GEMS brought about by NASA's TESS mission \citep{kanodia_searching_2024} surveys have found that the occurrence of giant planets decreases towards lower mass stars \citep{endl_exploring_2006, johnson_giant_2010, maldonado_connecting_2019, schlecker_rv-detected_2022, gan_occurrence_2023, bryant_occurrence_2023, mignon_radial_2025-1, glusman_searching_2026}. This low occurrence rate of giant planets around low-mass stars is expected from a planet formation perspective due to the low solid surface density and the long accretion times \citep{laughlin_core_2004, Burn2021,  ShibataHelled2025}. 

The existence of these GEMS likely necessitates anomalous disks that are suitable to formation by having longer lifetimes, more massive or more metal rich than average \citep{ShibataHelled2025}. Since GEMS are rather rare, it is quite likely that they are the outcome of massive ``outlier" protoplanetary disks that are not the median of the disk population \citep{kanodia_transiting_2024}. Another possibility is that the observed GEMS are the result of disk fragmentation \citep{boss_rapid_2006, BossKanodia2023}, although the efficiency of this mechanism around low-mass stars remains unconstrained.  

\begin{figure}[h]
    \centering
    \includegraphics[width=0.45\columnwidth]{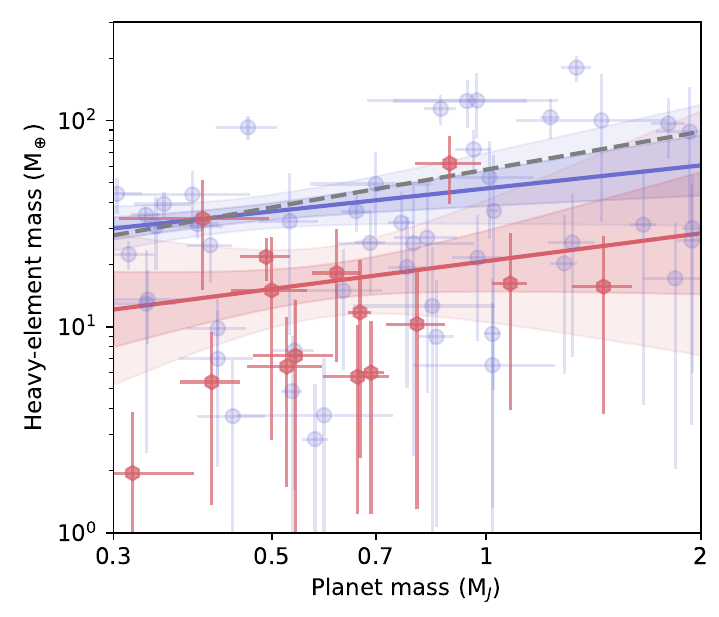}
    \caption{Inferred heavy-element mass (bottom) as a function of planet mass for $0.3 \leq M_p (M_J) \leq 2$. The solid lines show the fits constructed by Bayesian regression, and the shaded contours are the 1$\sigma$ and 2$\sigma$ uncertainties. Planets around FGK stars are depicted in blue, while those around M-dwarfs are in red. The grey dashed line shows the fit from \citet{2016ApJ...831...64T} for comparison. Figure from \citet{MH2025}.}
    \label{fig:metallicity_limited}
\end{figure}

In addition to their occurrence, estimates of the planetary bulk and atmospheric properties can help constrain planet formation theories \citep{madhusudhan_atmospheric_2017, knierim_convective_2024}. Studies of warm transiting giant planets have shown that the bulk densities seem to be agnostic of host stellar masses \citep{kanodia_transiting_2024, kanodia_searching_2025}. However, after accounting for their temperature and age, estimates of their bulk metallicities using interior models show hints of being lower for M-dwarf giant planets compared to those orbiting FGK stars \citep{MH2025}. 
The inferred heavy-element masses for the two samples are shown in Figure \ref{fig:metallicity_limited}. Both populations present an increasing trend for the heavy-element mass as a function of planetary mass. However, there is an apparent offset of the total heavy-elements mass for planets around M-stars in comparison to those around stars. This may suggest that GEMS are  overall metal-poor compared to their FGK counterparts. However, this result should be taken with caution since the  sample size is very small.  In addition, we note that \cite{Chachan2025} used a  sample of 147 warm giant planets, to re-assess  the planetary mass-metallicity relation and found no evidence of such a dependence. Therefore, empirically while it seems that the bulk density of GEMS is similar to that of FGK giant planets, it remains unknown whether the metallicity of GEMS is indeed different.  

It is also possible that there are  systematical differences in the interior temperature and/or ages of these GEMS (compared to the FGK sample). Another key observation is the atmospheric metallicity of giant planets. The Cycle 2 JWST program (GO 3171) --- \textit{Red Dwarfs and the Seven Giants: First Insights Into the Atmospheres of Giant Exoplanets around M dwarf Stars} \citep{kanodia_red_2023}, is starting to investigate this with a carefully chosen sample of seven GEMS. Initial results from this survey are finding planetary atmospheric metallicities that are 1 -- 2 dex substellar \citep{canas_gems_2026, ashtari_gems_2026, kanodia_gems_2026}.  
It is therefore interesting to explore  whether planet formation theory can explain this observation. Future atmospheric measurements, in particular those with ESA's Ariel mission can significantly increase the sample of GEMS to help infer trends and statistics in a more robust manner. In this paper, we briefly present the key arguments that explain why GEMS are rare, and why they are expected to have low bulk metallicities in comparison to giant planets orbiting sun-like stars. We argue that the GEMS population is key in revealing information on the conditions for giant planet formation by helping marginalize over the stochasticity in planet formation over the stellar mass axis.


\section{Formation of GEMS: Less heavy elements to begin with}\label{sec:isolation_masses}

\begin{table}
    \centering
    \begin{tabular}{c|l}
        \hline
        Symbol & Definition \\
        \hline
        $M_\star$ & Stellar mass \\
        $L_\star$ & Stellar luminosity \\
        $\dot{M}_\mathrm{disk}$ & Disk accretion rate \\
        $r$ & Distance from the central star \\
        $h$ & Disk scale height \\
        $\Sigma_\mathrm{g}$ & Gas surface density \\
        $\Sigma_\mathrm{s}$ & Planetesimal surface density \\
        $M_\mathrm{iso,plts}$ & Planetesimal isolation mass \\
        $M_\mathrm{iso,peb}$ & Pebble isolation mass \\
        $A$ & Scaling exponent for stellar luminosity \\
        $B$ & Scaling exponent for disk accretion rate \\
        $x$ & Scaling exponent for core formation location \\
        \hline
    \end{tabular}
    \caption{Definitions of the parameters used in this section.}
    \label{tab:params}
\end{table}

Typically, the heavy-element content of giant planets is fundamentally determined by the reservoir of solids available in the primordial protoplanetary disk. Observational surveys of young stellar objects across various star-forming regions (e.g., Lupus, Chamaeleon I) suggest a nearly linear, or perhaps super-linear, scaling between the dust mass of the disk and the mass of the host star \citep{Pascucci2016, Ansdell2017}. 
This suggests that giant planets around higher-mass stars have larger solid reservoirs, increasing their opportunity to accrete heavy elements. 

Before the onset of runaway gas accretion, the maximum core mass set by the available solid material is known as the isolation mass. In the core-accretion scenario, a planetary core first grows toward the isolation mass and subsequently begins to accrete a gaseous envelope \citep[e.g.,][]{Pollack1996}. 
The main phase of gas accretion begins after the core approaches isolation, because the critical core mass—the maximum core mass that can maintain a hydrostatic envelope without rapid contraction—decreases as the solid accretion rate declines \citep{Ikoma2000}. 
Whether a planetary core ultimately undergoes runaway gas accretion to become a gas giant depends on several factors, including its cooling efficiency and the lifetime of the protoplanetary disk \citep{ShibataHelled2025}, the population of observed giant planets is generally understood to originate from planetary cores that attained their isolation masses and entered runaway gas accretion prior to disk dispersal. 

For planetesimal accretion, the isolation mass is set by the available mass of planetesimals within the planetary feeding zone, and is given by: 
\begin{equation}\label{eq:Miso_plts}
    M_\mathrm{iso,plts} \propto \Sigma_\mathrm{s}^{3/2} M_{\star}^{-1/2} r^{3},
\end{equation}
where $\Sigma_\mathrm{s}$ is the surface density of planetesimals, $M_\star$ is the mass of the host star, and $r$ is the distance between the protoplanet and the star. 
In disks around low-mass stars, the lower absolute values of $\Sigma_s$ translate into smaller core masses and a reduced total heavy-element mass. 

In the pebble accretion paradigm, a critical threshold for planetary growth is the pebble isolation mass ($M_{\mathrm{iso,peb}}$). This represents the mass at which a growing protoplanet creates a pressure bump in the gas disk, trapping pebbles at its outer edge and effectively halting the accretion of solids \citep[e.g.,][]{Lambrechts2014}. The scaling of $M_{\text{iso,peb}}$ with stellar mass is vital for understanding the diversity of planetary compositions across different stellar types. The pebble isolation mass can be expressed as a function of the local disk properties, most notably the aspect ratio $h/r$ \citep{Brum2024}:
\begin{equation}
    M_{\mathrm{iso,peb}} \approx 20 
    \left( \frac{h/r}{0.05} \right)^3 
    \left( \frac{M_\star}{M_\odot} \right) 
    f(\alpha) M_{\oplus},
\end{equation}
where $f(\alpha)$ accounts for the effect of gas viscosity.

For discussing the dependence of the isolation mass on stellar mass, we follow the disk model used in \citet{Ida+2016}. The disk aspect ratio is determined by the balance between stellar irradiation and viscous heating, and scales as:
\begin{align}
    \frac{h}{r} \propto 
    \begin{cases}
        M_\star^{-7/20} \dot{M}_\mathrm{disk}^{1/5} r^{1/20}, 
        & \text{viscous heating}, \\
        L_\star^{1/7} M_\star^{-4/7} r^{2/7}, 
        & \text{irradiation heating},
    \end{cases}
\end{align}
where $\dot{M}_\mathrm{disk}$ is the disk accretion rate, and $L_\star$ is the luminosity of the central star. 
The transition radius between viscous and irradiation heating, $r_\mathrm{vis-irr}$, scales as $ r_\mathrm{vis-irr} \propto L_\star^{-20/33}\dot{M}_\mathrm{disk}^{28/33}M_\star^{31/33}$. For a solar-type star with $\dot{M}_\mathrm{disk}=10^{-8} M_\odot \mathrm{yr}^{-1}$, this transition occurs at $\sim 2~\mathrm{au}$. 
Assuming a steady-state disk, the gas surface density is given by:
\begin{align}
    \Sigma_\mathrm{g} \propto
    \begin{cases}
        M_\star^{1/5} \dot{M}_\mathrm{disk}^{3/5} r^{-3/5}, 
        & \text{viscous heating}, \\
        L_\star^{-2/7} M_\star^{9/14} \dot{M}_\mathrm{disk} r^{-15/14}, 
        & \text{irradiation heating}.
    \end{cases}
\end{align}

We can introduce scaling relations for the stellar luminosity,
$L_\star \propto M_\star^A$, and the disk accretion rate,
$\dot{M}_\mathrm{disk} \propto M_\star^B$. Combining these scaling laws with the assumption that the planetesimal surface density follows the gas surface density, the planetesimal isolation mass becomes: 
\begin{align}
    M_\mathrm{iso,plts} \propto 
    \begin{cases}
        M_\star^{9B/10-1/5} r^{21/10}, 
        & \text{viscous heating}, \\
        M_\star^{-3A/7 + 3B/2 + 13/28} r^{39/28}, 
        & \text{irradiation heating}.
    \end{cases}
\end{align}
Similarly, the pebble isolation mass is: 
\begin{align}
    M_\mathrm{iso,peb} \propto 
    \begin{cases}
        M_\star^{3B/5-1/20} r^{3/20}, 
        & \text{viscous heating}, \\
        M_\star^{3A/7 - 5/7} r^{6/7}, 
        & \text{irradiation heating}.
    \end{cases}
\end{align}

For pre-main-sequence stars, the stellar luminosity scaling parameter is expected to lie in the range $A=1$--$3$. Observational surveys suggest $B=1$--$3$ for the scaling of the disk accretion rate \citep{Alcala2017}. For typical values of $A=2$ and $B=2$, the planetesimal isolation mass scales as $M_\star^{8/5}$ and $M_\star^{73/28}$ in the viscous-heating and irradiation-heating regimes, respectively. Similarly, the pebble isolation mass scales as $M_\star^{23/20}$ and $M_\star^{1/7}$ in the viscous-heating and irradiation-heating regimes, respectively. 

Also, the isolation mass depends on the distance from the central star. Both planetesimal and pebble accretion models suggest that planetary cores preferentially form outside the water ice line \citep{Emsenhuber2021, Savvidou2023}. In the planetesimal accretion scenario, the surface density of planetesimals increases beyond the ice line because of the condensation of water ice. In the pebble accretion scenario, sticky icy pebbles can grow larger, increasing the accretion efficiency. Therefore, the formation location of the planetary cores is expected to depend also on the disk temperature and stellar mass, suggesting that planetary cores form at smaller orbital distances around lower-mass stars. 

\begin{figure}
    \centering
    \includegraphics[width=0.9\linewidth]{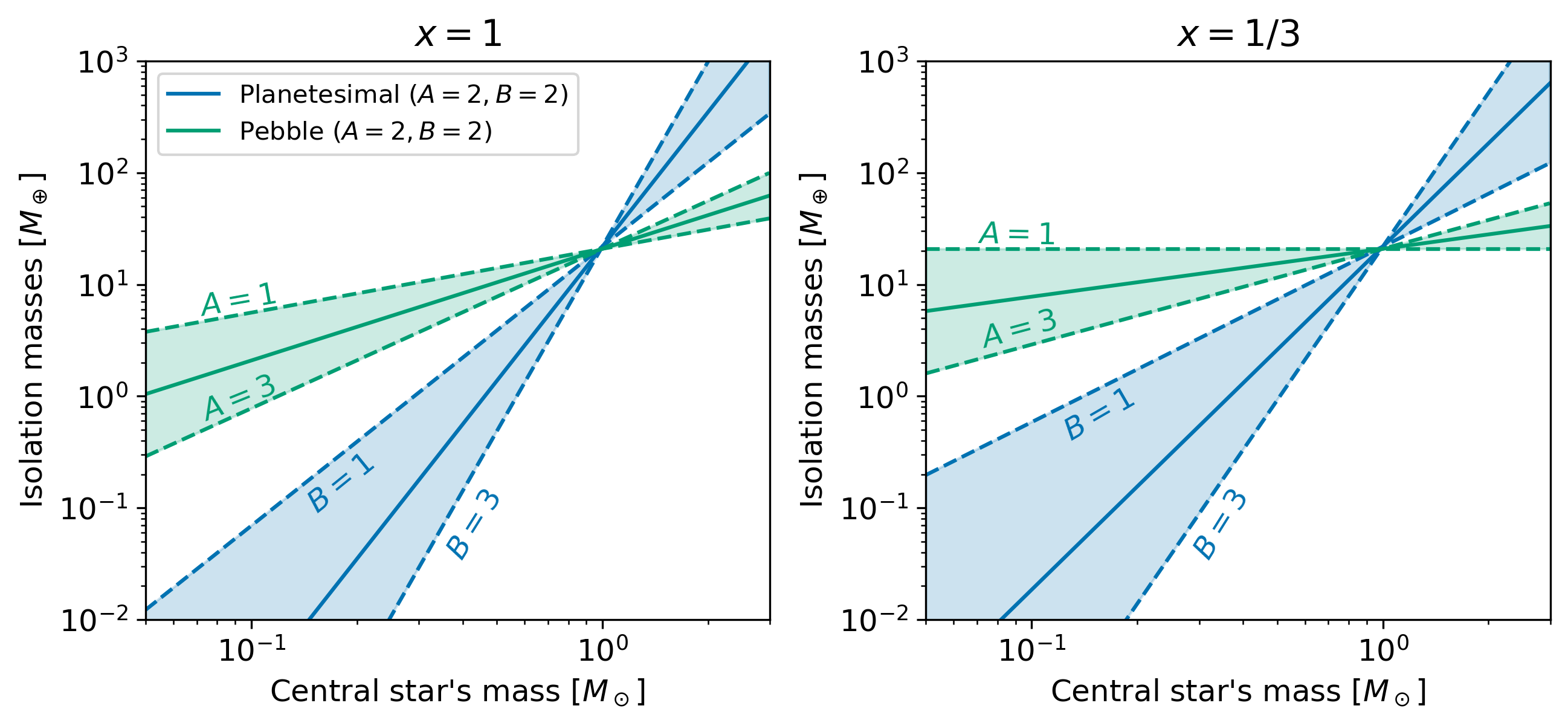}
    \caption{
    Isolation masses as functions of stellar mass. The blue and green lines show the planetesimal and pebble isolation masses, respectively. The solid lines correspond to the typical disk-scaling case with $A=2$ and $B=2$, while the dashed lines show cases with different scaling parameters. The left and right panels show results for different values of $x$, the scaling parameter that determines the core formation location (\autoref{eq:axi0_scaling}) We adopt $L_\star = L_\odot$ and $\dot{M}_\mathrm{disk} = 10^{-8} M_\odot \mathrm{yr}^{-1}$ for the solar-mass case. It is important to emphasize here that the range of isolation masses depicted here for a range of stellar and disk conditions will not all necessarily undergo runaway gaseous accretion. As mentioned earlier, the onset of this is dependent on not just the isolation mass (typically expected to be $\sim$ 10 $M_{\oplus}$), but also the grain opacity, cooling rate and formation timescales.
    }
    \label{fig:Ms_Miso}
\end{figure}

Figure \ref{fig:Ms_Miso} shows the isolation masses as a function of stellar masses. Here, we assumed that the formation location of planetary cores scales as: 
\begin{align}
    r_\mathrm{p,0} = 10~\mathrm{au}\left(\frac{M_\star}{M_\odot}\right)^{\expAxi}. \label{eq:axi0_scaling}
\end{align}
In the irradiation-dominated region, the disk's temperature  and also the location of ice lines scale  with the distance from the central star. Therefore, $x=1$ if planetary cores form in the same temperature region around different mass stars, and the corresponding isolation masses are shown in the left panel of  Fig.~\ref{fig:Ms_Miso}. 
We also adopt $L_\star = L_\odot$ and $\dot{M}_\mathrm{disk} = 10^{-8} M_\odot \mathrm{yr}^{-1}$ for the solar-mass case. For reference, we also adopt a solar disk metallicity of $0.015$. We first consider giant-planet formation in typical protoplanetary disks, with $A=2$ and $B=2$. In this case, both the planetesimal isolation mass (blue solid line) and the pebble isolation mass (green solid line) decrease strongly with decreasing stellar mass.

We also present the case with $x=1/3$ where planetary cores form in orbits that correspond to the  same orbital period around different mass stars. In this case, the dependence of the isolation masses on the stellar mass  weakens, but the trend that the isolation mass  decreases with decreasing stellar mass remains. 

The estimated occurrence rates of giant planets implies  that they preferentially form around stars with higher metallicities \citep{fulton_california_2021}. This suggests that giant-planet-hosting systems may represent a biased subset of protoplanetary disks, selected from systems with large solid (heavy-element) reservoirs. Consequently, the disk properties relevant to giant-planet formation may vary more weakly with stellar mass within this selected population than in the full population of protoplanetary disks. Under this assumption, we also plot the isolation masses with different scaling parameters, as indicated by the dashed lines in Fig.~\ref{fig:Ms_Miso}. 

Nevertheless, theoretical models generally predict lower isolation masses for planets forming around M dwarfs compared to those forming around sun-like stars, except the pebble isolation mass with $A=1$ and $x=1/3$. Therefore, the isolation mass tends to be lower around lower-mass stars, implying a smaller reservoir of solids available for planets forming around such stars and, consequently, a lower expected heavy-element content.

These scaling imply that even if a giant planet manages to trigger runaway gas accretion around a low-mass, which is clearly challenging given the longer accretion timescale \citep[see e.g.,][and references therein]{ShibataHelled2025} and lower surface densities \citep{Laughlin2004} it will necessarily do so in an environment where the supply of planetesimals or pebbles is exhausted more quickly. This lower isolation mass around low-mass stars creates a bottleneck for giant planet formation \citep{Brum2024,ShibataHelled2025}. If the isolation mass is reached before the core becomes massive enough to trigger runaway gas accretion, the planet may remain a sub-Neptune or an ice giant. This reinforces the idea that giant planets around M dwarfs are not only rarer, but depending on the disks in which they form may also be more metal-poor.

\section{Post-formation Enrichments via Planetesimal Accretion}

\subsection{Planetesimal accretion during runaway gas accretion}
Accretion of planetesimals by growing protoplanets in protoplanetary disks can significantly enrich the planets in heavy elements. Different from pebble accretion, the protoplanet could accrete additional planetesimals beyond the planetesimal isolation mass after entering the runaway gas-accretion phase, because rapid gas accretion further expands the planetary feeding zone \citep{Zhou2007, Shiraishi2008, Shibata2019}. Thus, planetesimals remaining in the vicinity of the protoplanet can continue to be accreted. 

Numerous theoretical studies have investigated this process in the context of exoplanetary giant planets, estimating both the total mass and composition of the accreted planetesimals 
\citep{Shibata2020, Hands2021, Turrini2021, Shibata2022b, Knierim2022, Pacetti2022}. However, most of these studies focus on gas giant planets forming around Sun-like stars. As shown by the analytical prescription of \citet{Shibata2022b}, the planetesimal accretion onto a migrating protoplanet depends on the stellar mass and the properties of the surrounding disk. To quantify how stellar properties affect planetesimal accretion, we perform a suite of $N$-body simulations around stars spanning a range of masses.

We investigate how many planetesimals are accreted by a growing and migrating protoplanet. We extend the numerical framework of \citet{Shibata2020} to account for variations in stellar mass and luminosity by adopting the scaling disk model by \citet{Ida+2016}, introduced in Section~\ref{sec:isolation_masses}. We adopt $A=2$ and treat $B$ as a free parameter because it controls the planetesimal surface-density profile. We assume that the protoplanet undergoes runaway gas accretion from $10 M_\oplus$ to a Jupiter-mass, coupled with inward type-II migration. We adopt the $\alpha$-viscosity prescription \citep{Shakura+1973} and set the viscosity parameter to $10^{-3}$. The resulting migration timescale is $\sim 1 \times 10^6$ yr at 10 au around a solar-mass star. The migration timescale is longer around lower-mass stars and at larger orbital distances. The initial semi-major axis of the protoplanet is modeled with Eq.~\ref{eq:axi0_scaling}.
Here, we perform two sets of simulations with $\expAxi=1$,  and $1/3$. We terminate the simulations once the planet reaches $0.5$~au, as interior to this location planetesimal accretion becomes negligible. Planetesimals are subject to gravitational scattering by the growing protoplanet as well as aerodynamic gas drag from the surrounding protoplanetary disk. If the planets originate from outer orbital region, planetesimal accretion happens in a wider disk area. Also, the surface density of planetesimals is higher around higher mass stars because of the larger mass budget of heavy elements. Therefore, the planet around higher mass stars could accrete more planetesimals. Further details of our numerical model are described in the Appendix.

\begin{figure}
    \centering
    \includegraphics[width=0.6\linewidth]{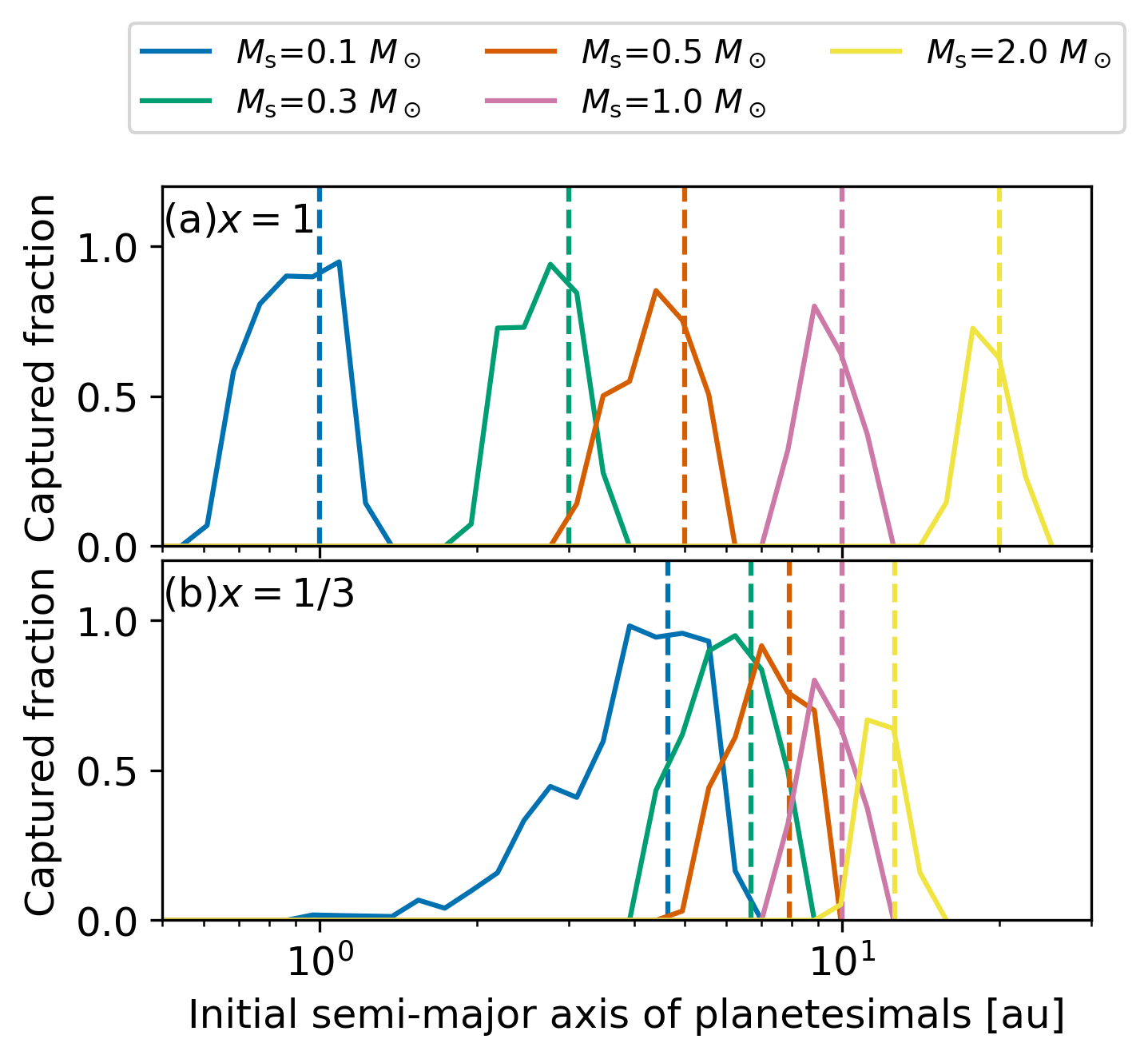}
    \caption{Accretion efficiency of planetesimals as a function of the initial semi-major axis. We shows the cases with typical disk scaling parameter $B=2$. Each panel shows the results obtained using a different scaling law for the initial position of a growing protoplanet. We divide the range $a_\mathrm{init}=0.1$--$30$~au into 50 logarithmically spaced bins. Different colors correspond to simulations with different stellar masses. The vertical dashed lines indicate the initial orbits of planetary cores.} 
    \label{fig:M_acc_migration}
\end{figure}

Figure~\ref{fig:M_acc_migration} shows the inferred accretion efficiency of planetesimals as a function of their initial semi-major axis for different stellar masses. Here, we show the cases with the typical scaling parameter $B=2$. The accretion efficiency is defined as the number of captured planetesimals divided by the total number of planetesimals initially distributed in each bin. The different panels correspond to different assumed values of $x$. Most planetesimals are accreted near the orbit where the planetary core forms, although additional planetesimals from the inner disk are also accreted as the planet migrates toward the disk inner edge.

\begin{figure}
    \centering
    \includegraphics[width=0.6\linewidth]{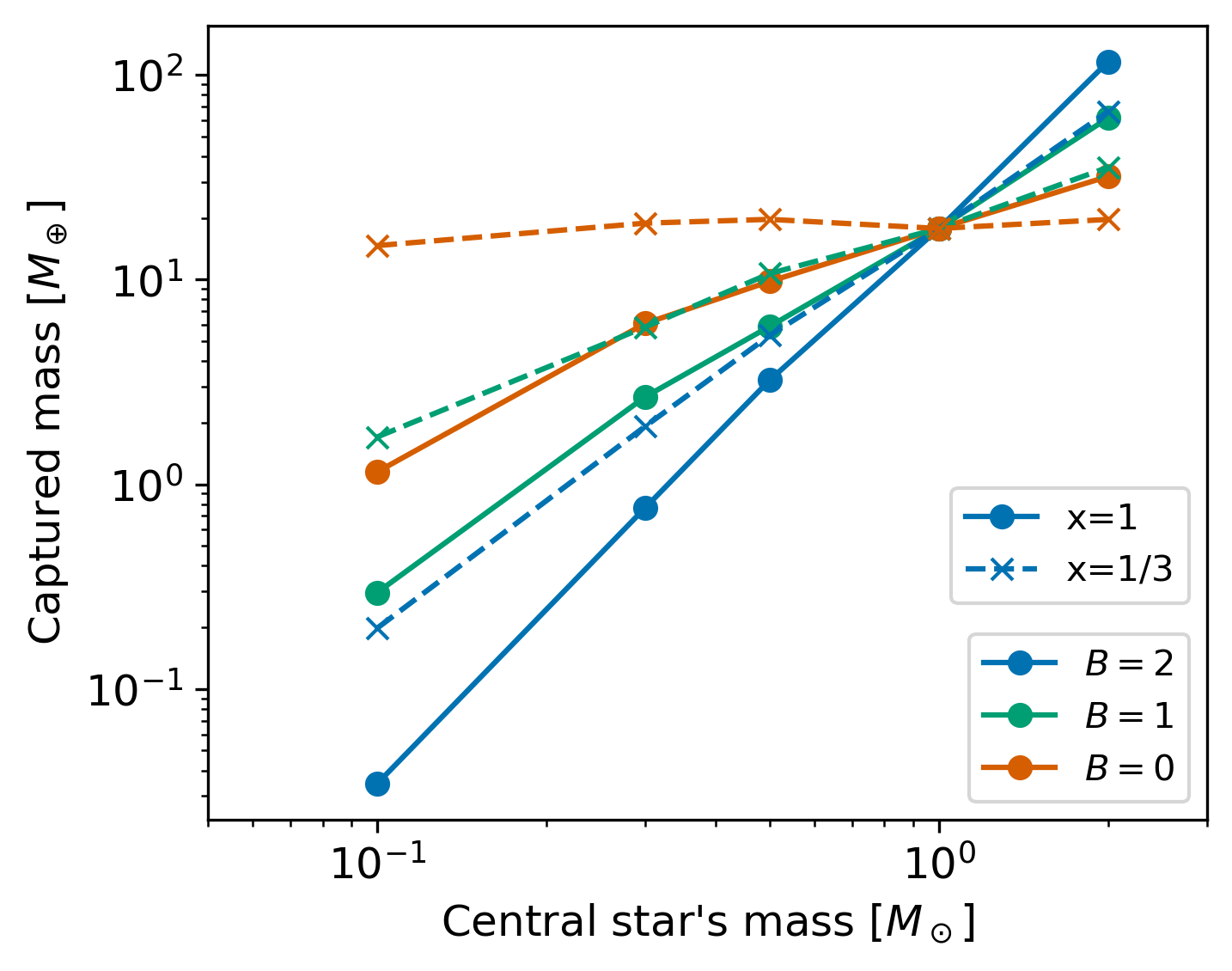}
    \caption{Captured planetesimal mass as a function of the central star's mass. The blue, green, and orange plots show the results for $B = 2$, $1$, and $0$, respectively. The circle and cross markers denote for $x=1$ and $1/3$, respectively.}
    \label{fig:Ms_Mcap}
\end{figure}

Although the accretion efficiency depends only weakly on the initial orbit of the planet, planets accrete more heavy elements when they migrate from outer regions, because these regions contain larger reservoirs of planetesimals. We summarize the results of our numerical experiments in Figure~\ref{fig:Ms_Mcap}. We plot the inferred accreted heavy-element mass as a function of stellar mass for each combination of $B$ and $x$. In most cases, our numerical results show a positive correlation between the accreted heavy-element mass and stellar mass. This trend mainly reflects the larger solid reservoirs available around higher-mass stars. The exception is the case with $B=0$ and $x=1/3$. Only in this extreme case, giant planets accrete similar masses of heavy elements around stars of different masses.

We note that the planetesimal accretion rate depends on both the planetary migration timescale \citep{Shibata2020,Shibata2022b} and the protoplanet growth timescale \citep{Shibata2019}. For example, more rapid planetary migration can increase the efficiency of planetesimal accretion. \citet{Shibata2020} found that a planet migrating ten times faster can accrete approximately twice as many planetesimals. Such migration rates may be realized in protoplanetary disks with higher viscosities.  Therefore, the inferred heavy-element mass accreted by a planet depends on the details of the formation model, as well as on the disk structure, evolution, and assumed viscosity. Although our main conclusion---that planets formed in the outer disk accrete more heavy elements---is likely robust, further studies of planetesimal accretion around stars of different masses are essential for understanding the compositions of gas giant planets.

\subsection{Heavy-element enrichment after disk dissipation} 
The final heavy-element budget of a giant planet can also be affected by accretion of small objects after the gaseous disk had dissipated \citep{Seligman2022, Rodet2023, Shibata2024}.  
\citet{Wyatt2017} investigated the dynamics of the scattering of planetesimals or planetary embryos forming around stars with different masses. In particular, they investigated the planet mass versus semimajor axis parameter space and identified the dominant outcome of scattering interactions with a planet of given parameters. This provides information on when planetesimals are expected to be accreted or ejected. It was found that accretion of solids is more efficient around  more massive, Sun-like stars.  In contrast, around low-mass stars (M-dwarfs), even a moderately massive protoplanet can easily enter the scattering-dominated limit, where planetesimals are preferentially ejected or scattered out of the feeding zone before they can be accreted. This dynamical filtering suggests that, independent of disk mass, giant planets around low-mass stars are intrinsically less efficient at capturing heavy elements than those around Solar-type hosts.



Finally, we also note that the total heavy-element enrichment in giant planets can be linked to the planetary formation location and migration history. 
If a giant planet forms at larger semi-major axes and subsequently undergoes orbital migration, it must pass through regions where scattering is the dominant dynamical outcome. During this transit, the planet is expected to "clear" its path by scattering planetesimals rather than accreting them. This disparity is more significant around low-mass stars, where the transition to the scattering regime occurs at lower planetary masses. Around sun-like stars, planetesimals can be accreted onto a migrating giant planets as demonstrated above and as found in previous studies \citep{Shibata2020, Hands2021, Shibata2022b, Knierim2022}.

\section{Connection to Ariel}

In the previous sections we discuss the formation and planet evolution conditions and assumptions that could contribute to potentially lower bulk metallicities for GEMS, compared to FGK giant planets. Empirically, this quantity is hard to constrain for exoplanets (and perhaps even Jupiter/Saturn), given the degeneracies caused by potential compositional gradients, equations of state for H/He under the pressure-temperature conditions of planetary interiors as well as uncertainty associated with cooling timescales. This is partially ameliorated by observing a carefully selected sample of planets \citep{batalha_importance_2023} with similar ages and temperatures, and using interior models such as \texttt{planetsynth} \citep{muller_synthetic_2021}, \texttt{GASTLI} \citep{acuna_gastli_2024-1} or \texttt{APPLE }\citep{sur_apple_2024}, which are further constrained by the measured atmospheric metallicity (typically obtained for transiting planets through transmission or emission spectroscopy). In this context, ESA's upcoming Ariel mission \citep{tinetti_science_2016} can help with its large wavelength coverage (0.5 - 7.8 $\mu$m, simultaneously) and dedicated atmospheres focus by  providing a large sample of GEMS with measured atmospheric metallicities and energy budgets.

\subsection{Atmospheric Metallicity}
Measurements of atmospheric metallicity, particularly for planets around M-dwarfs can be fraught due to effects of stellar contamination, as has been seen with the JWST program studying seven GEMS \citep{kanodia_red_2023, canas_gems_2026}. The effects of stellar contamination are caused by stellar heterogeneities on the surface of the star \citep{rackham_transit_2018}, such as star spots and faculae. For M-dwarfs, the star spots are not enough to thermally dissociate water, the presence of which in star spots can bias planetary atmospheric measurements \citep[e.g., ][]{moran_high_2023}. While this bias is largely due to inaccurate M-dwarf stellar line lists and models, it can be considerably mitigated by transmission observations in the Rayleigh-Jeans tail of the stellar blackbody, i.e., the mid-infrared \citep{seager_why_2024}. Ariel's wavelength coverage extending to 7.8 $\mu$m should allow for more accurate constraints on stellar contamination than the NIR, which can then be used to provide more robust planetary atmospheric composition estimates.


\subsection{Energy Budgets}

\begin{figure}
    \centering
    \includegraphics[width=0.8\linewidth]{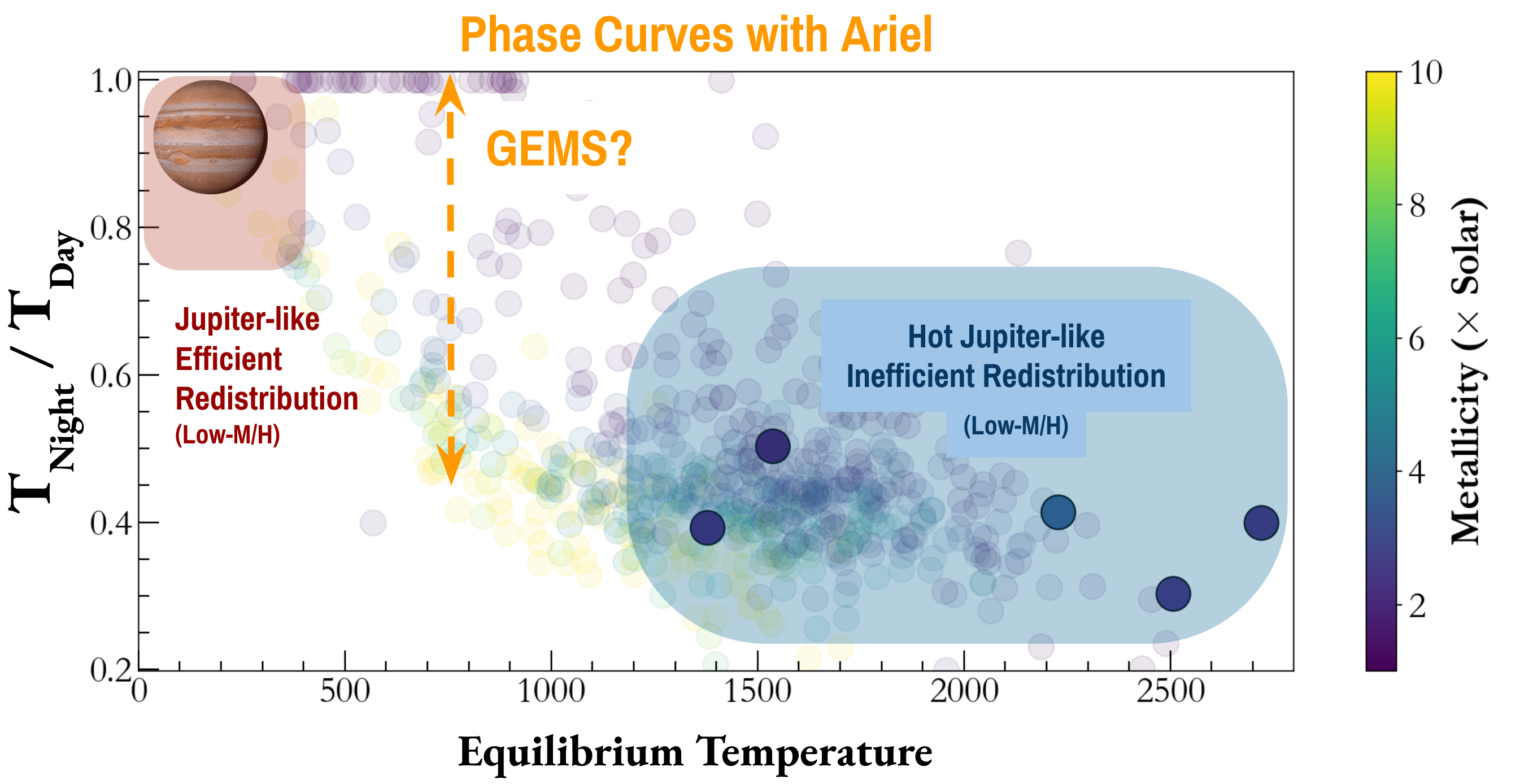}
    \caption{Planets with JWST phase curves (solid), show inefficient redistribution (blue) that is discrepant from the almost perfect heat redistribution seen for Jupiter (red). GEMS (orange) bridge the gap between Jupiter and hot Jupiters. Measuring their heat redistribution across this parameter space will also help constrain the factors that influence giant planet climates and energy budgets that can then used to improve bulk metallicity estimates for a large sample of GEMS.}
    \label{fig:ArielPhaseCurves}
\end{figure}

Observations of the variation in planetary flux as a function of the orbital phase can reveal information about a planet's global atmospheric chemistry and climate, including the temperature distribution, efficiency of day/night heat transport, reflectivity (albedos), and the response of a planet to stellar tides \citep{stevenson_thermal_2014, parmentier_exoplanet_2018}. GEMS with their larger flux ratios and relative planet radii (relative to warm Jupiters) are particularly favourable for such measurements \citep{teinturier_warm_2024-1}, and sit in a temperature regime between our Jupiter ($\sim 150 K$) and currently studied hot Jupiters ($\gtrsim$ 1000 K), as shown in \autoref{fig:ArielPhaseCurves}. About 10\% of ARIEL's science time is expected to be devoted to phase curves \citep{charnay_survey_2022}, testing heat distribution, global climate models, and cloud properties for warm gas giants. Therefore, a phase curve survey of a representative sample of GEMS, could help constrain the energy budgets for GEMS, which can then be used to further marginalize over some of the degeneracies in planetary interior models. 

We explore the potential of Ariel to observe phase curves of GEMS by applying the methods of \citet{charnay_ariel_pc} to the latest Ariel target list \citep{edwards_ariel_2022}. Briefly, we limit the sample to known GEMS and then compute the expected phase curve amplitude. We then used ArielRad \citep{mugnai_arielrad} to determine the precision of the Ariel observations and thus the signal-to-noise ratio that Ariel should achieve in a single phase curve. We do this for two scenarios. Firstly, we assess the available targets if we assume Ariel's Tier 1 binning which essentially leads to 7 photometric phase curves. These which could be used to build temperature maps of the planet, constraining the day-night contrast and thus the efficiency of heat redistribution. We find that there are five planets for which a single phase curve would lead to an SNR$>$7 on the amplitude. The second scenario we consider are spectroscopic phase curves at the Tier 2 resolution of Ariel. Ariel's Tier 2 are sensitive to molecular features, allowing for the chemistry (metallicity, C/O, etc.) to be constrained \citep[e.g.,][]{changeat_alfnoor}. In this case, they would allow us to probe changes in chemistry with phase. We find two targets, TOI-3714b and TOI-3235b, that would be ideal for study in this manner. We note that, in both cases, these numbers are relatively small. However, the main limitation is not the sensitivity of Ariel's infrared instrumentation but of the fine guidance sensor (FGS). Ariel has two FGS bands: FGS1 (0.6-0.8 $\mu$m) and FGS2 (0.8-1.1 $\mu$m). For M-dwarfs, these naturally receive a limited amount of flux, particularly given the 10\,Hz readout. If alternations were made to the FGS operation, such as reading-out at a slower rate, the number of suitable targets would drastically increase. Without the current FGS limitation, we find over 20 targets that could achieve high-quality Tier 1 photometric phase curves in a single visit. How many of these could actually be realised would depend upon the FGS mitigation strategies that were employed. The use of this is speculative, and would require significant work by the project team, but we highlight the significant gain in the number of targets as a potential way of motivating the investigation of these options.

Therefore,  Ariel observations of a \textit{well-chosen} sample of GEMS will augment existing efforts with JWST \citep{kanodia_red_2023, canas_gems_2026} to measure their atmospheric metallicities and subsequently infer their bulk metallicities. By increasing the sample size that currently shows hints of a sharp divergence in atmospheric metallicity between M-dwarf and FGK giant planets, although their bulk properties seem more agnostic of the hots star mass.

\section{Discussion and Conclusions}
Our current understanding of the formation and composition of GEMS  is still limited, primarily due to the relatively small number of well-characterized planets in this category. 
To advance our knowledge, more observational data is crucial. This includes the detection of a larger sample of giant planets around M dwarfs with precise measurements of their masses and radii. Detailed characterization of their atmospheric compositions using techniques like transmission and emission spectroscopy is also essential to provide further constraints on their elemental abundances. Furthermore, obtaining more accurate determinations of the fundamental parameters of M dwarf host stars, such as their mass, radius, age, and metallicity, is critical to refining our understanding of the planetary systems they host. 
\par

On the theoretical side, our model does not consider the accretion of heavy elements through disk gas accretion. Volatiles evaporated from pebbles as they cross their ice lines can enhance the local metallicity of the disk gas \citep{Booth2017,Booth2019}. By accreting this metal-enriched gas, giant planets can acquire up to several tens of Earth masses of heavy elements around G-type stars \citep{Schneider2021a,Schneider2021b}. However, it is still unknown how this accretion channel scales with stellar mass. Since the observed metallicity of GEMS should reflect the combined contribution of heavy elements accreted through both solid accretion and gas accretion, future formation models should include both processes, properly coupled with disk chemical evolution.

It is also crucial to note here that while the analysis shown her assumes nominal initial disk conditions, it is quite likely that these planets do not form in average disks given their rarity \citep{kanodia_transiting_2024}. The average M-dwarf protoplanetary disk likely forms protoplanets too slowly, or with isolation masses that are too low to undergo runaway gaseous accretion (\autoref{fig:Ms_Miso}), hence explaining the low occurrence of these objects \citep{bryant_occurrence_2023-1, glusman_searching_2026}. However the observed distribution of M-dwarf protoplanetary disk masses also includes anomalous disks that can be orders of magnitude more massive than the average \citep{Manara2023}. GEMS likely require initial disk masses that are consistent with $\gtrsim$ 90th percentile in mass distribution of the disk sample. In this scenario the conditions should enable their formation, but \textit{might} still imprint onto their present-day bulk metallicities. Therefore future work incorporating a population synthesis Monte-Carlo approach with reasonable priors for initial conditions could simultaneously explain the rarity of these objects and provide a tangible range bulk metallicities (or bulk densities) to compare with the observed sample \citep{kanodia_transiting_2024, MuellerHelled2025}.

Ariel is expected to play a pivotal role in providing crucial new data for characterizing the atmospheres and for  constraining the bulk compositions of giant planets around M dwarfs. Continued theoretical modeling and simulations are also vital for exploring the vast parameter space of planet formation around M dwarfs and for providing a framework to interpret the growing body of observational data. Finally, further research into the connection between the atmospheres and interiors of these planets is needed to better understand how atmospheric observations relate to their overall bulk composition.

While the fundamental mechanisms of planet formation may be universal, the resulting planetary populations exhibit diversity across different stellar environments. Future research, leveraging larger observational samples and advanced instruments, alongside continued theoretical investigations, are essential to solidify the conclusions and further refine our understanding of the fascinating realm of exoplanetary systems around low-mass stars.

\begin{acknowledgments}
 We thank Simon Müller and Lorenzo Peerani for valuable discussions. 
\end{acknowledgments}

\begin{contribution}
All authors contributed equally to the paper.


\end{contribution}

%

\appendix

\section{Models and Settings of Planetesimal Accretion Simulation}

We adopt the planetesimal-accretion framework developed by \citet{Shibata2020}. We assume that a growing and migrating protoplanet accretes planetesimals distributed throughout the protoplanetary disk. Each simulation includes 3,360 super-particles representing planetesimals. By performing N-body simulations, we investigate how the mass of accreted planetesimals varies with the mass of the central star.

Our disk model is based on \citet{Ida+2016}, in which the stellar luminosity and the disk accretion rate scale with stellar mass. We assume that the protoplanet enters the runaway gas accretion phase at $t = 1$ Myr. The protoplanet then grows exponentially from $10 M_\oplus$ to one Jupiter mass, following the model of \citet{Turrini2021}, with an adopted growth timescale of $10^5$ yr.

The growing protoplanet migrates inward with classical type-II migration: once a planet opens a deep gap, the inner and outer disk become decoupled, and the planet migrates inward toward the central star at approximately the viscous accretion speed of the gas. Many hydrodynamical simulations indicate that planetary gaps can be substantially shallower and may not fully separate the protoplanetary disk \citep{Kanagawa+2018}. In this case, the migration regime differs from the classical type-II picture and the migration rate is reduced. Such ``slow'' type-II migration is broadly consistent with the observed occurrence of cold Jupiters, which must avoid large-scale inward migration during their formation. In contrast, hot and warm Jupiters---the focus of this study---are difficult to produce under slow type-II migration alone. One possible solution is wind-driven migration, which emerges in hydrodynamical simulations based on the disk wind model \citep{Lega2021}. In this framework, the system transitions to the classical type-II regime when the planetary torque becomes strong enough to impede the wind-driven accretion flow across the gap. Motivated by this picture, we implement the type-II migration timescale as
\begin{align}
    \tau_{\rm mig,II} = \frac{2 r^2}{3 \nu}
    \left( 1 + \frac{M_{\rm p}}{2 \pi \Sigma_{\rm gas} r^2} \right).
\end{align}

\bibliography{sample7, MyLibrary}{}

@article{acuna_gastli_2024-1,
  title = {{{GASTLI}}. {{An}} Open-Source Coupled Interior-Atmosphere Model to Unveil Gas-Giant Composition},
  author = {Acu{\~n}a, L. and Kreidberg, L. and Zhai, M. and Molli{\`e}re, P.},
  year = 2024,
  month = aug,
  journal = {Astronomy and Astrophysics},
  volume = {688},
  pages = {A60},
  publisher = {EDP},
  issn = {0004-6361},
  doi = {10.1051/0004-6361/202450559},
  urldate = {2025-10-28}
}

@article{ashtari_gems_2026,
  title = {{{GEMS JWST}}: {{HATS-75}} b {{A Giant Planet}} with a {{Subsolar Metallicity Atmosphere Orbiting}} an {{M Dwarf}}},
  shorttitle = {{{GEMS JWST}}},
  author = {Ashtari, Reza and {Lustig-Yaeger}, Jacob and {Libby-Roberts}, Jessica and M{\"u}ller, Simon and Kanodia, Shubham and Stevenson, Kevin B. and Ca{\~n}as, Caleb I. and Guzm{\'a}n Caloca, Giannina and Wallack, Nicole L. and Delamer, Megan and Piette, Anjali A. A. and Mahadevan, Suvrath and Czekala, Ian and Han, Te and Helled, Ravit},
  year = 2026,
  month = apr,
  journal = {AJ},
  volume = {171},
  number = {5},
  pages = {294},
  publisher = {The American Astronomical Society},
  issn = {1538-3881},
  doi = {10.3847/1538-3881/ae552c},
  urldate = {2026-04-17},
  langid = {english}
}

@article{batalha_importance_2023,
  title = {Importance of {{Sample Selection}} in {{Exoplanet-atmosphere Population Studies}}},
  author = {Batalha, Natasha E. and Wolfgang, Angie and Teske, Johanna and Alam, Munazza K. and Alderson, Lili and Batalha, Natalie M. and {L{\'o}pez-Morales}, Mercedes and Wakeford, Hannah R.},
  year = 2023,
  month = jan,
  journal = {The Astronomical Journal},
  volume = {165},
  pages = {14},
  publisher = {IOP},
  issn = {0004-6256},
  doi = {10.3847/1538-3881/ac9f45},
  urldate = {2025-10-28}
}

@article{boss_rapid_2006,
  title = {Rapid {{Formation}} of {{Gas Giant Planets}} around {{M Dwarf Stars}}},
  author = {Boss, Alan P.},
  year = 2006,
  month = may,
  journal = {The Astrophysical Journal},
  volume = {643},
  pages = {501--508},
  issn = {0004-637X},
  doi = {10.1086/501522},
  urldate = {2021-04-18}
}

@misc{bryant_occurrence_2023,
  title = {The Occurrence Rate of Giant Planets Orbiting Low-Mass Stars with {{TESS}}},
  author = {Bryant, Edward M and Bayliss, Daniel and Van Eylen, Vincent},
  year = 2023,
  month = mar,
  journal = {arXiv e-prints},
  urldate = {2023-03-02}
}

@article{bryant_occurrence_2023-1,
  title = {The Occurrence Rate of Giant Planets Orbiting Low-Mass Stars with {{TESS}}},
  author = {Bryant, Edward M. and Bayliss, Daniel and Van Eylen, Vincent},
  year = 2023,
  month = may,
  journal = {Monthly Notices of the Royal Astronomical Society},
  volume = {521},
  pages = {3663--3681},
  publisher = {OUP},
  issn = {0035-8711},
  doi = {10.1093/mnras/stad626},
  urldate = {2025-10-27}
}

@article{canas_gems_2026,
  title = {{{GEMS JWST}}: {{Transmission Spectroscopy}} of {{TOI-5205b Reveals Significant Stellar Contamination}} and a {{Metal-poor Atmosphere}}},
  shorttitle = {{{GEMS JWST}}},
  author = {Ca{\~n}as, Caleb I. and {Lustig-Yaeger}, Jacob and Tsai, Shang-Min and M{\"u}ller, Simon and Helled, Ravit and Kanodia, Shubham and Louie, Dana R. and Guzm{\'a}n Caloca, Giannina and Gao, Peter and {Libby-Roberts}, Jessica and {Hardegree-Ullman}, Kevin K. and Col{\'o}n, Knicole D. and Czekala, Ian and Delamer, Megan and Han, Te and Lin, Andrea S.J. and Mahadevan, Suvrath and May, Erin M. and Ninan, Joe P. and Piette, Anjali A. A. and Stef{\'a}nsson, Gu{\dh}mundur and Stevenson, Kevin B. and Teske, Johanna and Wallack, Nicole L.},
  year = 2026,
  month = mar,
  journal = {AJ},
  volume = {171},
  number = {4},
  pages = {260},
  publisher = {The American Astronomical Society},
  issn = {1538-3881},
  doi = {10.3847/1538-3881/ae4976},
  urldate = {2026-04-01},
  langid = {english}
}

@article{charnay_survey_2022,
  title = {A Survey of Exoplanet Phase Curves with {{Ariel}}},
  author = {Charnay, Benjamin and Mendon{\c c}a, Jo{\~a}o M. and Kreidberg, Laura and Cowan, Nicolas B. and Taylor, Jake and Bell, Taylor J. and Demangeon, Olivier and Edwards, Billy and Haswell, Carole A. and Morello, Giuseppe and Mugnai, Lorenzo V. and Pascale, Enzo and Tinetti, Giovanna and Tremblin, Pascal and Zellem, Robert T.},
  year = 2022,
  month = apr,
  journal = {Exp Astron},
  volume = {53},
  number = {2},
  pages = {417--446},
  issn = {1572-9508},
  doi = {10.1007/s10686-021-09715-x},
  urldate = {2026-07-09},
  langid = {english}
}

@article{dressing_occurrence_2015,
  title = {The {{Occurrence}} of {{Potentially Habitable Planets Orbiting M Dwarfs Estimated}} from the {{Full Kepler Dataset}} and an {{Empirical Measurement}} of the {{Detection Sensitivity}}},
  author = {Dressing, Courtney D. and Charbonneau, David},
  year = 2015,
  month = jul,
  journal = {The Astrophysical Journal},
  volume = {807},
  pages = {45},
  issn = {0004-637X},
  doi = {10.1088/0004-637X/807/1/45},
  urldate = {2019-01-18}
}

@misc{edwards_ariel_2022,
  title = {The {{Ariel Target List}}: {{The Impact}} of {{TESS}} and the {{Potential}} for {{Characterising Multiple Planets Within}} a {{System}}},
  shorttitle = {The {{Ariel Target List}}},
  author = {Edwards, Billy and Tinetti, Giovanna},
  year = 2022,
  month = jul,
  number = {arXiv:2205.05073},
  eprint = {2205.05073},
  publisher = {arXiv},
  doi = {10.48550/arXiv.2205.05073},
  urldate = {2024-11-19},
  archiveprefix = {arXiv}
}

@article{endl_exploring_2006,
  title = {Exploring the {{Frequency}} of {{Close-in Jovian Planets}} around {{M Dwarfs}}},
  author = {Endl, Michael and Cochran, William D. and K{\"u}rster, Martin and Paulson, Diane B. and Wittenmyer, Robert A. and MacQueen, Phillip J. and Tull, Robert G.},
  year = 2006,
  month = sep,
  journal = {The Astrophysical Journal},
  volume = {649},
  pages = {436--443},
  issn = {0004-637X},
  doi = {10.1086/506465},
  urldate = {2021-04-16}
}

@article{fulton_california_2021,
  title = {California {{Legacy Survey II}}. {{Occurrence}} of {{Giant Planets Beyond}} the {{Ice}} Line},
  author = {Fulton, Benjamin J. and Rosenthal, Lee J. and Hirsch, Lea A. and Isaacson, Howard and Howard, Andrew W. and Dedrick, Cayla M. and Sherstyuk, Ilya A. and Blunt, Sarah C. and Petigura, Erik A. and Knutson, Heather A. and Behmard, Aida and Chontos, Ashley and Crepp, Justin R. and Crossfield, Ian J. M. and Dalba, Paul A. and Fischer, Debra A. and Henry, Gregory W. and Kane, Stephen R. and Kosiarek, Molly and Marcy, Geoffrey W.},
  year = 2021,
  month = may,
  journal = {arXiv:2105.11584 [astro-ph]},
  eprint = {2105.11584},
  primaryclass = {astro-ph},
  urldate = {2021-05-26},
  archiveprefix = {arXiv}
}

@article{gan_occurrence_2023,
  title = {Occurrence {{Rate}} of {{Hot Jupiters Around Early-type M Dwarfs Based}} on {{Transiting Exoplanet Survey Satellite Data}}},
  author = {Gan, Tianjun and Wang, Sharon X. and Wang, Songhu and Mao, Shude and Huang, Chelsea X. and Collins, Karen A. and Stassun, Keivan G. and Shporer, Avi and Zhu, Wei and Ricker, George R. and Vanderspek, Roland and Latham, David W. and Seager, Sara and Winn, Joshua N. and Jenkins, Jon M. and Barkaoui, Khalid and Belinski, Alexander A. and Ciardi, David R. and Evans, Phil and Girardin, Eric and Maslennikova, Nataliia A. and Mazeh, Tsevi and Panahi, Aviad and Pozuelos, Francisco J. and Radford, Don J. and Schwarz, Richard P. and Twicken, Joseph D. and W{\"u}nsche, Ana{\"e}l and Zucker, Shay},
  year = 2023,
  month = jan,
  journal = {AJ},
  volume = {165},
  pages = {17},
  issn = {0004-6256},
  doi = {10.3847/1538-3881/ac9b12},
  urldate = {2023-03-09}
}

@article{glusman_searching_2026,
  title = {Searching for {{GEMS}}: {{The Occurrence}} of {{Giant Planets Orbiting M Dwarfs}} within 100 Pc},
  shorttitle = {Searching for {{GEMS}}},
  author = {Glusman, Rowen I. and Ca{\~n}as, Caleb I. and Kanodia, Shubham and Han, Te and Fernandes, Rachel B. and Stef{\'a}nsson, Gu{\dj}mundur and Ford, Eric B. and Maney, Marissa and Monson, Andrew and Hotnisky, Andrew and Mahadevan, Suvrath and Rodruck, Michael and Ment, Kristo and McWilliam, Andrew and Cochran, William D. and Col{\'o}n, Knicole D. and Giovinazzi, Mark R. and {Alvarado-Montes}, Jaime A. and Bender, Chad F. and Blake, Cullen H. and Boone, Alexandra and Diddams, Scott A. and Gupta, Arvind F. and Halverson, Samuel and Krolikowski, Daniel and Lin, Andrea S. J. and Ninan, Joe P. and Robertson, Paul and Roy, Arpita and Schwab, Christian and Terrien, Ryan and Teske, Johanna and Wright, Jason T.},
  year = 2026,
  month = mar,
  journal = {The Astronomical Journal},
  volume = {171},
  pages = {146},
  publisher = {IOP},
  issn = {0004-6256},
  doi = {10.3847/1538-3881/ae2fbb},
  urldate = {2026-02-21}
}

@article{hardegree-ullman_kepler_2019,
  title = {Kepler {{Planet Occurrence Rates}} for {{Mid-type M Dwarfs}} as a {{Function}} of {{Spectral Type}}},
  author = {{Hardegree-Ullman}, Kevin K. and Cushing, Michael C. and Muirhead, Philip S. and Christiansen, Jessie L.},
  year = 2019,
  month = aug,
  journal = {The Astronomical Journal},
  volume = {158},
  pages = {75},
  issn = {0004-6256},
  doi = {10.3847/1538-3881/ab21d2},
  urldate = {2020-04-26}
}

@article{henry_solar_1994,
  title = {The Solar Neighborhood, 1: {{Standard}} Spectral Types ({{K5-M8}}) for Northern Dwarfs within Eight Parsecs},
  shorttitle = {The Solar Neighborhood, 1},
  author = {Henry, Todd J. and Kirkpatrick, J. Davy and Simons, Douglas A.},
  year = 1994,
  month = oct,
  journal = {The Astronomical Journal},
  volume = {108},
  pages = {1437--1444},
  doi = {10.1086/117167},
  urldate = {2020-08-15}
}

@article{johnson_giant_2010,
  title = {Giant {{Planet Occurrence}} in the {{Stellar Mass-Metallicity Plane}}},
  author = {Johnson, John Asher and Aller, Kimberly M. and Howard, Andrew W. and Crepp, Justin R.},
  year = 2010,
  month = aug,
  journal = {Publications of the Astronomical Society of the Pacific},
  volume = {122},
  pages = {905},
  issn = {0004-6280},
  doi = {10.1086/655775},
  urldate = {2021-09-17}
}

@misc{kanodia_gems_2026,
  title = {{{GEMS JWST}}: {{A}} Sub-{{Solar}} Metallicity Atmosphere for Giant Planet {{TOI-5293Ab}} Orbiting a Rapidly Changing {{M-dwarf}}},
  shorttitle = {{{GEMS JWST}}},
  author = {Kanodia, Shubham and Ca{\~n}as, Caleb I. and {Lustig-Yaeger}, Jacob and Guzm{\'a}n Caloca, Giannina and Wallack, Nicole L. and M{\"u}ller, Simon and Helled, Ravit and Col{\'o}n, Knicole D. and Czekala, Ian and Delamer, Megan and Han, Te and {Libby-Roberts}, Jessica and Piette, Anjali A. A. and Stevenson, Kevin B. and Stefansson, Gudmundur and Teske, Johanna},
  year = 2026,
  month = mar,
  publisher = {arXiv},
  doi = {10.48550/arXiv.2603.16464},
  urldate = {2026-03-20}
}

@article{kanodia_red_2023,
  title = {Red {{Dwarfs}} and the {{Seven Giants}}: {{First Insights}} into the {{Atmospheres}} of {{Giant Exoplanets}} around {{M-dwarf Stars}}},
  shorttitle = {Red {{Dwarfs}} and the {{Seven Giants}}},
  author = {Kanodia, Shubham and Canas, Caleb and {Libby-Roberts}, Jessica and Colon, Knicole and Czekala, Ian and Gao, Peter and Helled, Ravit and Lin, Andrea and Mahadevan, Suvrath and Mueller, Simon and Ninan, Joe Philip and Piette, Anjali A. A. and Stefansson, Gudmundur and Stevenson, Kevin and Teske, Johanna and Tsai, Shang-Min and Wallack, Nicole Lisa},
  year = 2023,
  month = may,
  journal = {JWST Proposal. Cycle 2},
  pages = {3171},
  urldate = {2023-10-04}
}

@article{kanodia_searching_2024,
  title = {Searching for {{Giant Exoplanets}} around {{M-dwarf Stars}} ({{GEMS}}) {{I}}: {{Survey Motivation}}},
  shorttitle = {Searching for {{Giant Exoplanets}} around {{M-dwarf Stars}} ({{GEMS}}) {{I}}},
  author = {Kanodia, Shubham and Ca{\~n}as, Caleb I. and Mahadevan, Suvrath and Ford, Eric B. and Helled, Ravit and Anderson, Dana E. and Boss, Alan and Cochran, William D. and Delamer, Megan and Han, Te and {Libby-Roberts}, Jessica E. and Lin, Andrea S. J. and M{\"u}ller, Simon and Robertson, Paul and Stef{\'a}nsson, Gumundur and Teske, Johanna},
  year = 2024,
  month = apr,
  journal = {AJ},
  volume = {167},
  pages = {161},
  issn = {0004-6256},
  doi = {10.3847/1538-3881/ad27cb},
  urldate = {2024-03-31}
}

@article{kanodia_searching_2025,
  title = {Searching for {{GEMS}}: {{TOI-7149}} b, an {{Inflated Giant Planet Causing}} a 12\% {{Transit}} of a {{Fully Convective M-dwarf}}},
  shorttitle = {Searching for {{GEMS}}},
  author = {Kanodia, Shubham and Ca{\~n}as, Caleb I. and Mahadevan, Suvrath and Lin, Andrea S. J. and Kobulnicky, Henry A. and Karfs, Ian and Birkholz, Alexina and Monson, Andrew and Gupta, Arvind F. and Everett, Mark and Rodruck, Michael and Glusman, Rowen I. and Han, Te and Cochran, William D. and Bender, Chad F. and Diddams, Scott A. and Krolikowski, Daniel and Halverson, Samuel and {Libby-Roberts}, Jessica and Ninan, Joe P. and Robertson, Paul and Roy, Arpita and Schwab, Christian and Stef{\'a}nsson, Gu{\~d}mundur},
  year = 2025,
  month = oct,
  journal = {AJ},
  volume = {170},
  pages = {203},
  publisher = {IOP},
  issn = {0004-6256},
  doi = {10.3847/1538-3881/adf6db},
  urldate = {2025-10-27}
}

@article{kanodia_transiting_2024,
  title = {Transiting {{Jupiters}} around {{M Dwarfs Have Similar Masses}} to {{FGK Warm Jupiters}}},
  author = {Kanodia, Shubham},
  year = 2024,
  month = dec,
  journal = {ApJ},
  volume = {978},
  number = {1},
  pages = {97},
  publisher = {The American Astronomical Society},
  issn = {0004-637X},
  doi = {10.3847/1538-4357/ad9823},
  urldate = {2024-12-27},
  langid = {english}
}

@misc{knierim_convective_2024,
  title = {Convective Mixing in Giant Planets and the Connection to Atmospheric Measurements},
  author = {Knierim, Henrik and Helled, Ravit},
  year = 2024,
  month = jul,
  journal = {arXiv e-prints},
  urldate = {2024-07-15}
}

@article{laughlin_core_2004,
  title = {The {{Core Accretion Model Predicts Few Jovian-Mass Planets Orbiting Red Dwarfs}}},
  author = {Laughlin, Gregory and Bodenheimer, Peter and Adams, Fred C.},
  year = 2004,
  month = sep,
  journal = {ApJL},
  volume = {612},
  pages = {L73-L76},
  issn = {0004-637X},
  doi = {10.1086/424384},
  urldate = {2021-02-18}
}

@article{madhusudhan_atmospheric_2017,
  title = {Atmospheric Signatures of Giant Exoplanet Formation by Pebble Accretion},
  author = {Madhusudhan, Nikku and Bitsch, Bertram and Johansen, Anders and Eriksson, Linn},
  year = 2017,
  month = aug,
  journal = {Monthly Notices of the Royal Astronomical Society},
  volume = {469},
  pages = {4102--4115},
  issn = {0035-8711},
  doi = {10.1093/mnras/stx1139},
  urldate = {2020-04-27}
}

@article{maldonado_connecting_2019,
  title = {Connecting Substellar and Stellar Formation: The Role of the Host Star's Metallicity},
  shorttitle = {Connecting Substellar and Stellar Formation},
  author = {Maldonado, J. and Villaver, E. and Eiroa, C. and Micela, G.},
  year = 2019,
  month = apr,
  journal = {Astronomy \&amp; Astrophysics, Volume 624, id.A94, {$<$}NUMPAGES{$>$}7{$<$}/NUMPAGES{$>$} pp.},
  volume = {624},
  pages = {A94},
  issn = {0004-6361},
  doi = {10.1051/0004-6361/201833827},
  urldate = {2021-07-28},
  langid = {english}
}

@article{ment_occurrence_2023,
  title = {The {{Occurrence Rate}} of {{Terrestrial Planets Orbiting Nearby Mid-to-late M Dwarfs}} from {{TESS Sectors}} 1-42},
  author = {Ment, Kristo and Charbonneau, David},
  year = 2023,
  month = jun,
  journal = {The Astronomical Journal},
  volume = {165},
  pages = {265},
  issn = {0004-6256},
  doi = {10.3847/1538-3881/acd175},
  urldate = {2024-03-05}
}

@article{mignon_radial_2025-1,
  title = {Radial Velocity Homogeneous Analysis of {{M}} Dwarfs Observed with {{HARPS}}: {{II}}. {{Detection}} Limits and Planetary Occurrence Statistics},
  shorttitle = {Radial Velocity Homogeneous Analysis of {{M}} Dwarfs Observed with {{HARPS}}},
  author = {Mignon, L. and Delfosse, X. and Meunier, N. and Chaverot, G. and Burn, R. and Bonfils, X. and Bouchy, F. and {Astudillo-Defru}, N. and Lo Curto, G. and Gaisne, G. and Udry, S. and Forveille, T. and Segransan, D. and Lovis, C. and Santos, N. C. and Mayor, M.},
  year = 2025,
  month = aug,
  journal = {Astronomy and Astrophysics},
  volume = {700},
  pages = {A146},
  publisher = {EDP},
  issn = {0004-6361},
  doi = {10.1051/0004-6361/202451142},
  urldate = {2025-10-27}
}

@article{moran_high_2023,
  title = {High {{Tide}} or {{Riptide}} on the {{Cosmic Shoreline}}? {{A Water-rich Atmosphere}} or {{Stellar Contamination}} for the {{Warm Super-Earth GJ}} 486b from {{JWST Observations}}},
  shorttitle = {High {{Tide}} or {{Riptide}} on the {{Cosmic Shoreline}}?},
  author = {Moran, Sarah E. and Stevenson, Kevin B. and Sing, David K. and MacDonald, Ryan J. and Kirk, James and {Lustig-Yaeger}, Jacob and Peacock, Sarah and Mayorga, L. C. and Bennett, Katherine A. and {L{\'o}pez-Morales}, Mercedes and May, E. M. and Rustamkulov, Zafar and Valenti, Jeff A. and Adams Redai, J{\'e}a I. and Alam, Munazza K. and Batalha, Natasha E. and Fu, Guangwei and {Gonzalez-Quiles}, Junellie and Highland, Alicia N. and Kruse, Ethan and Lothringer, Joshua D. and Ortiz Ceballos, Kevin N. and Sotzen, Kristin S. and Wakeford, Hannah R.},
  year = 2023,
  month = may,
  journal = {The Astrophysical Journal},
  volume = {948},
  pages = {L11},
  publisher = {IOP},
  issn = {0004-637X},
  doi = {10.3847/2041-8213/accb9c},
  urldate = {2026-07-02}
}

@article{muller_synthetic_2021,
  title = {Synthetic Evolution Tracks of Giant Planets},
  author = {M{\"u}ller, Simon and Helled, Ravit},
  year = 2021,
  month = oct,
  journal = {MNRAS},
  volume = {507},
  pages = {2094--2102},
  issn = {0035-8711},
  doi = {10.1093/mnras/stab2250},
  urldate = {2022-11-16}
}

@incollection{parmentier_exoplanet_2018,
  title = {Exoplanet {{Phase Curves}}: {{Observations}} and {{Theory}}},
  shorttitle = {Exoplanet {{Phase Curves}}},
  booktitle = {Handbook of {{Exoplanets}}},
  author = {Parmentier, Vivien and Crossfield, Ian J. M.},
  year = 2018,
  pages = {1419--1440},
  publisher = {Springer, Cham},
  doi = {10.1007/978-3-319-55333-7_116},
  urldate = {2026-07-09},
  isbn = {978-3-319-55333-7},
  langid = {english}
}

@article{rackham_transit_2018,
  title = {The {{Transit Light Source Effect}}: {{False Spectral Features}} and {{Incorrect Densities}} for {{M-dwarf Transiting Planets}}},
  shorttitle = {The {{Transit Light Source Effect}}},
  author = {Rackham, Benjamin V. and Apai, D{\'a}niel and Giampapa, Mark S.},
  year = 2018,
  month = feb,
  journal = {The Astrophysical Journal},
  volume = {853},
  pages = {122},
  issn = {0004-637X},
  doi = {10.3847/1538-4357/aaa08c},
  urldate = {2023-07-27}
}

@article{reyle_10_2021,
  title = {The 10 Parsec Sample in the {{Gaia}} Era},
  author = {Reyl{\'e}, C. and Jardine, K. and Fouqu{\'e}, P. and Caballero, J. A. and Smart, R. L. and Sozzetti, A.},
  year = 2021,
  month = jun,
  journal = {A\&A},
  volume = {650},
  pages = {A201},
  issn = {0004-6361},
  doi = {10.1051/0004-6361/202140985},
  urldate = {2021-09-09},
  langid = {english}
}

@article{schlecker_rv-detected_2022,
  title = {{{RV-detected}} Planets around {{M}} Dwarfs: {{Challenges}} for Core Accretion Models},
  shorttitle = {{{RV-detected}} Planets around {{M}} Dwarfs},
  author = {Schlecker, M. and Burn, R. and Sabotta, S. and Seifert, A. and Henning, Th and Emsenhuber, A. and Mordasini, C. and Reffert, S. and Shan, Y. and Klahr, H.},
  year = 2022,
  month = aug,
  journal = {Astronomy and Astrophysics},
  volume = {664},
  pages = {A180},
  issn = {0004-6361},
  doi = {10.1051/0004-6361/202142543},
  urldate = {2023-01-09},
  langid = {english}
}

@article{seager_why_2024,
  title = {Why {{Observations}} at {{Mid-infrared Wavelengths Partially Mitigate M Dwarf Star Host Stellar Activity Contamination}} in {{Exoplanet Transmission Spectroscopy}}},
  author = {Seager, Sara and Shapiro, Alexander I.},
  year = 2024,
  month = jul,
  journal = {ApJ},
  volume = {970},
  number = {2},
  pages = {155},
  publisher = {The American Astronomical Society},
  issn = {0004-637X},
  doi = {10.3847/1538-4357/ad509a},
  urldate = {2024-07-30},
  langid = {english}
}

@article{stevenson_thermal_2014,
  title = {Thermal Structure of an Exoplanet Atmosphere from Phase-Resolved Emission Spectroscopy},
  author = {Stevenson, Kevin B. and D{\'e}sert, Jean-Michel and Line, Michael R. and Bean, Jacob L. and Fortney, Jonathan J. and Showman, Adam P. and Kataria, Tiffany and Kreidberg, Laura and McCullough, Peter R. and Henry, Gregory W. and Charbonneau, David and Burrows, Adam and Seager, Sara and Madhusudhan, Nikku and Williamson, Michael H. and Homeier, Derek},
  year = 2014,
  month = nov,
  journal = {Science},
  volume = {346},
  number = {6211},
  pages = {838--841},
  publisher = {American Association for the Advancement of Science},
  doi = {10.1126/science.1256758},
  urldate = {2024-09-30}
}

@misc{sur_apple_2024,
  title = {{{APPLE}}: {{An Evolution Code}} for {{Modeling Giant Planets}}},
  shorttitle = {{{APPLE}}},
  author = {Sur, Ankan and Su, Yubo and Arevalo, Roberto Tejada and Chen, Yi-Xian and Burrows, Adam},
  year = 2024,
  month = apr,
  number = {arXiv:2404.14483},
  eprint = {2404.14483},
  primaryclass = {astro-ph},
  publisher = {arXiv},
  urldate = {2024-04-24},
  archiveprefix = {arXiv}
}

@article{teinturier_warm_2024-1,
  title = {Warm {{Jupiters}} around {{M}} Dwarfs Are Great Opportunities for Extensive Chemical, Cloud, and Haze Characterisation with {{JWST}}},
  author = {Teinturier, L. and Ducrot, E. and Charnay, B.},
  year = 2024,
  month = oct,
  journal = {Astronomy and Astrophysics},
  volume = {690},
  pages = {A380},
  publisher = {EDP},
  issn = {0004-6361},
  doi = {10.1051/0004-6361/202450761},
  urldate = {2025-10-28}
}

@inproceedings{tinetti_science_2016,
  title = {The Science of {{ARIEL}} (Atmospheric Remote-Sensing Infrared Exoplanet Large-Survey)},
  booktitle = {Space Telescopes and Instrumentation 2016: {{Optical}}, Infrared, and Millimeter Wave},
  author = {Tinetti, G. and Drossart, P. and Eccleston, P. and Hartogh, P. and Heske, A. and Leconte, J. and Micela, G. and Ollivier, M. and Pilbratt, G. and Puig, L. and Turrini, D. and Vandenbussche, B. and Wolkenberg, P. and Pascale, E. and Beaulieu, J. -P. and G{\"u}del, M. and Min, M. and Rataj, M. and Ray, T. and Ribas, I. and Barstow, J. and Bowles, N. and Coustenis, A. and {Coud{\'e} du Foresto}, V. and Decin, L. and Encrenaz, T. and Forget, F. and Friswell, M. and Griffin, M. and Lagage, P. O. and Malaguti, P. and Moneti, A. and Morales, J. C. and Pace, E. and Rocchetto, M. and Sarkar, S. and Selsis, F. and Taylor, W. and Tennyson, J. and Venot, O. and Waldmann, I. P. and Wright, G. and Zingales, T. and {Zapatero-Osorio}, M. R.},
  editor = {MacEwen, Howard A. and Fazio, Giovanni G. and Lystrup, Makenzie and Batalha, Natalie and Siegler, Nicholas and Tong, Edward C.},
  year = 2016,
  month = jul,
  series = {Society of Photo-Optical Instrumentation Engineers ({{SPIE}}) Conference Series},
  volume = {9904},
  pages = {99041X},
  doi = {10.1117/12.2232370}
}

@ARTICLE{Ikoma2000,
  title     = "Formation of giant planets: Dependences on core accretion rate
               and grain opacity",
  author    = "Ikoma, Masahiro and Nakazawa, Kiyoshi and Emori, Hiroyuki",
  journal   = "Astrophys. J.",
  publisher = "American Astronomical Society",
  volume    =  537,
  number    =  2,
  pages     = "1013--1025",
  month     =  jul,
  year      =  2000,
  language  = "en"
}

@article{Ansdell2017,
  author  = {{Ansdell}, M. and {Williams}, J.~P. and {Manara}, C.~F. and others},
  title   = {{An ALMA Survey of Lupus Protoplanetary Disks. II. Continuum Sizes and Radial Variations}},
  journal = {The Astronomical Journal},
  year    = {2017},
  volume  = {153},
  number  = {5},
  pages   = {240},
  doi     = {10.3847/1538-3881/aa69c0}
}

@article{Laughlin2004,
  author  = {{Laughlin}, G. and {Bodenheimer}, P. and {Adams}, F.~C.},
  title   = {{The Core Accretion Model Predicts Few Jovian-Mass Planets around Low-Mass Stars}},
  journal = {The Astrophysical Journal Letters},
  year    = {2004},
  volume  = {612},
  number  = {1},
  pages   = {L73--L76},
  doi     = {10.1086/424638}
}

@article{Brum2024,
  author  = {{Brum}, A.~J. and {Venturini}, J. and {Helled}, R.},
  title   = {{The effect of stellar mass on the pebble isolation mass}},
  journal = {Astronomy \& Astrophysics},
  year    = {2024},
  note    = {(Check latest volume/page info as this is a recent/upcoming study)}
}

@article{JWST2026,
  author  = {{Redfield}, S. and others},
  title   = {{New Constraints on the M Dwarf Cosmic Shoreline: Insights from the JWST Rocky Worlds Program}},
  journal = {The Astrophysical Journal Letters},
  year    = {2026},
  volume  = {in press}
}

@ARTICLE{Zhou2007,
  title     = "Planetesimal Accretion onto Growing {Proto–Gas} Giant Planets",
  author    = "Zhou, Ji‐lin and Lin, Douglas N C",
  journal   = "Astrophys. J.",
  publisher = "IOP Publishing",
  volume    =  666,
  number    =  1,
  pages     = "447--465",
  month     =  sep,
  year      =  2007
}

@ARTICLE{Shiraishi2008,
  title     = "Infall of Planetesimals onto Growing Giant Planets: Onset of
               Runaway Gas Accretion and Metallicity of Their Gas Envelopes",
  author    = "Shiraishi, Masakazu and Ida, Shigeru",
  journal   = "Astrophys. J.",
  publisher = "IOP Publishing",
  volume    =  684,
  number    =  2,
  pages     = "1416--1426",
  month     =  sep,
  year      =  2008
}

@article{Pascucci2016,
  author  = {{Pascucci}, I. and {Testi}, L. and {Herczeg}, G.~J. and others},
  title   = {{A Steeper rather than Flatter Disk Mass-Stellar Mass Relation}},
  journal = {The Astrophysical Journal},
  year    = {2016},
  volume  = {831},
  number  = {2},
  pages   = {125},
  doi     = {10.3847/0004-637X/831/2/125}
}

@article{Pollack1996,
  author  = {{Pollack}, J.~B. and {Hubickyj}, O. and {Bodenheimer}, P. and others},
  title   = {{Formation of the Giant Planets by Concurrent Accretion of Solids and Gas}},
  journal = {Icarus},
  year    = {1996},
  volume  = {124},
  number  = {1},
  pages   = {62--85},
  doi     = {10.1006/icar.1996.0188}
}

@article{Greene2023,
  author  = {{Greene}, T.~P. and others},
  title   = {{Thermal Emission from the Earth-sized Exoplanet TRAPPIST-1 b using JWST}},
  journal = {Nature},
  year    = {2023},
  volume  = {618},
  pages   = {39--42}
}

@ARTICLE{Chachan2025,
       author = {{Chachan}, Yayaati and {Fortney}, Jonathan J. and {Ohno}, Kazumasa and {Thorngren}, Daniel and {Murray-Clay}, Ruth},
        title = "{Revising the Giant Planet Mass─Metallicity Relation: Deciphering the Formation Sequence of Giant Planets}",
      journal = {\apj},
         year = 2025,
        month = nov,
       volume = {994},
       number = {1},
          eid = {43},
        pages = {43},
          doi = {10.3847/1538-4357/ae0cbf},
archivePrefix = {arXiv},
       eprint = {2509.20428},
 primaryClass = {astro-ph.EP},
       adsurl = {https://ui.adsabs.harvard.edu/abs/2025ApJ...994...43C}
}

@article{Lambrechts2014,
  author  = {{Lambrechts}, M. and {Johansen}, A. and {Moriguchi}, A.},
  title   = {{Separating gas-giant and ice-giant planets by pebble accretion}},
  journal = {Astronomy \& Astrophysics},
  year    = {2014},
  volume  = {572},
  pages   = {A35},
  doi     = {10.1051/0004-6361/201423814}
}

@article{ShibataHelled2025,
  author       = {Shibata, Sho and Helled, Ravit},
  title        = {Giant planet formation via pebble accretion across different stellar masses},
  journal      = {Astronomy \& Astrophysics},
  year         = {2025},
  volume       = {700},
  pages        = {A224},
  doi          = {10.1051/0004-6361/202555787},
  url          = {https://doi.org/10.1051/0004-6361/202555787}
}

@article{MuellerHelled2025,
  author       = {Müller, Simon and Helled, Ravit},
  title        = {The bulk metallicity of giant planets around M stars},
  journal      = {Astronomy \& Astrophysics},
  year         = {2025},
  volume       = {693},
  pages        = {L4},
  doi          = {10.1051/0004-6361/202452442},
  url          = {https://doi.org/10.1051/0004-6361/202452442}
}

@article{BossKanodia2023,
  author       = {Boss, Alan P. and Kanodia, Shubham},
  title        = {Forming Gas Giants Around a Range of Protostellar M-dwarfs by Gas Disk Gravitational Instability},
  journal      = {The Astrophysical Journal},
  year         = {2023},
  volume       = {956},
  number       = {1},
  pages        = {4},
  doi          = {10.3847/1538-4357/ac058a},
  url          = {https://doi.org/10.3847/1538-4357/ac058a}
}

@article{Burn2021,
  author       = {Burn, Remo and Schlecker, M. and Mordasini, C. and Emsenhuber, A. and Alibert, Y. and Henning, T. and Klahr, H. and Benz, W.},
  title        = {The New Generation Planetary Population Synthesis (NGPPS). IV. Planetary systems around low-mass stars},
  journal      = {Astronomy \& Astrophysics},
  year         = {2021},
  volume       = {656},
  pages        = {A72},
  doi          = {10.1051/0004-6361/202140390},
  url          = {https://doi.org/10.1051/0004-6361/202140390}
}

@ARTICLE{MH2025,
       author = {{M{\"u}ller}, Simon and {Helled}, Ravit},
        title = "{The bulk metallicity of giant planets around M stars}",
      journal = {\aap},
         year = 2025,
        month = jan,
       volume = {693},
          eid = {L4},
        pages = {L4},
          doi = {10.1051/0004-6361/202452442},
archivePrefix = {arXiv},
       eprint = {2411.16197},
 primaryClass = {astro-ph.EP},
       adsurl = {https://ui.adsabs.harvard.edu/abs/2025A&A...693L...4M}
}

@ARTICLE{Wyatt2017,
       author = {{Wyatt}, M.~C. and {Bonsor}, A. and {Jackson}, A.~P. and {Marino}, S. and {Shannon}, A.},
        title = "{How to design a planetary system for different scattering outcomes: giant impact sweet spot, maximizing exocomets, scattered discs}",
      journal = {\mnras},
         year = 2017,
        month = jan,
       volume = {464},
       number = {3},
        pages = {3385-3407},
          doi = {10.1093/mnras/stw2633},
archivePrefix = {arXiv},
       eprint = {1610.00714},
 primaryClass = {astro-ph.EP},
       adsurl = {https://ui.adsabs.harvard.edu/abs/2017MNRAS.464.3385W}
}

@ARTICLE{mugnai_arielrad,
       author = {{Mugnai}, Lorenzo V. and {Pascale}, Enzo and {Edwards}, Billy and {Papageorgiou}, Andreas and {Sarkar}, Subhajit},
        title = "{ArielRad: the Ariel radiometric model}",
      journal = {Experimental Astronomy},
         year = 2020,
        month = oct,
       volume = {50},
       number = {2-3},
        pages = {303-328},
          doi = {10.1007/s10686-020-09676-7},
archivePrefix = {arXiv},
       eprint = {2009.07824},
 primaryClass = {astro-ph.IM},
       adsurl = {https://ui.adsabs.harvard.edu/abs/2020ExA....50..303M}
}

@ARTICLE{charnay_ariel_pc,
       author = {{Charnay}, Benjamin and {Mendon{\c{c}}a}, Jo{\~a}o M. and {Kreidberg}, Laura and {Cowan}, Nicolas B. and {Taylor}, Jake and {Bell}, Taylor J. and {Demangeon}, Olivier and {Edwards}, Billy and {Haswell}, Carole A. and {Morello}, Giuseppe and {Mugnai}, Lorenzo V. and {Pascale}, Enzo and {Tinetti}, Giovanna and {Tremblin}, Pascal and {Zellem}, Robert T.},
        title = "{A survey of exoplanet phase curves with Ariel}",
      journal = {Experimental Astronomy},
         year = 2022,
        month = apr,
       volume = {53},
       number = {2},
        pages = {417-446},
          doi = {10.1007/s10686-021-09715-x},
archivePrefix = {arXiv},
       eprint = {2102.06523},
 primaryClass = {astro-ph.EP},
       adsurl = {https://ui.adsabs.harvard.edu/abs/2022ExA....53..417C}
}

@ARTICLE{changeat_alfnoor,
       author = {{Changeat}, Q. and {Al-Refaie}, A. and {Mugnai}, L.~V. and {Edwards}, B. and {Waldmann}, I.~P. and {Pascale}, E. and {Tinetti}, G.},
        title = "{Alfnoor: A Retrieval Simulation of the Ariel Target List}",
      journal = {\aj},
         year = 2020,
        month = aug,
       volume = {160},
       number = {2},
          eid = {80},
        pages = {80},
          doi = {10.3847/1538-3881/ab9a53},
archivePrefix = {arXiv},
       eprint = {2003.01839},
 primaryClass = {astro-ph.EP},
       adsurl = {https://ui.adsabs.harvard.edu/abs/2020AJ....160...80C}
}

@ARTICLE{Shakura+1973,
   author = {{Shakura}, N.~I. and {Sunyaev}, R.~A.},
    title = "{Black holes in binary systems. Observational appearance.}",
  journal = {Astronomy {\&} Astrophysics},
     year = 1973,
   volume = 24,
    pages = {337-355},
   adsurl = {http://ads.nao.ac.jp/abs/1973A\%26A....24..337S}
}

@ARTICLE{Ida+2016,
  title     = "The radial dependence of pebble accretion rates: A source of
               diversity in planetary systems: {I}. Analytical formulation",
  author    = "Ida, S and Guillot, T and Morbidelli, A",
  journal   = "Astron. Astrophys.",
  publisher = "EDP Sciences",
  volume    =  591,
  pages     = "A72",
  month     =  jul,
  year      =  2016
}

@ARTICLE{Kanagawa+2018,
  title     = "Radial Migration of Gap-opening Planets in Protoplanetary Disks.
               {I}. The Case of a Single Planet",
  author    = "Kanagawa, Kazuhiro D and Tanaka, Hidekazu and Szuszkiewicz, Ewa",
  journal   = "Astrophys. J.",
  publisher = "IOP Publishing",
  volume    =  861,
  number    =  2,
  pages     =  140,
  month     =  jul,
  year      =  2018
}

@ARTICLE{Shibata2019,
  title    = "Capture of solids by growing proto-gas giants: effects of gap
              formation and supply limited growth",
  author   = "Shibata, Sho and Ikoma, Masahiro",
  journal  = "Mon. Not. R. Astron. Soc.",
  volume   =  487,
  number   =  4,
  pages    = "4510--4524",
  month    =  aug,
  year     =  2019
}

@ARTICLE{Shibata2020,
  title     = "The origin of the high metallicity of close-in giant exoplanets -
               Combined effects of resonant and aerodynamic shepherding",
  author    = "Shibata, Sho and Helled, Ravit and Ikoma, Masahiro",
  journal   = "Astron. Astrophys. Suppl. Ser.",
  publisher = "EDP Sciences",
  volume    =  633,
  pages     = "A33",
  month     =  jan,
  year      =  2020,
  language  = "en"
}

@ARTICLE{Lega2021,
  title     = "Migration of Jupiter mass planets in discs with laminar accretion
               flows",
  author    = "Lega, E and Morbidelli, A and Nelson, R P and Ramos, X S and
               Crida, A and Bethune, A W and Batygin, K",
  journal   = "Astron. Astrophys.",
  publisher = "EDP Sciences",
  month     =  nov,
  year      =  2021
}

@ARTICLE{Hands2021,
  title     = "Super stellar abundances of alkali metals suggest significant
               migration for hot Jupiters",
  author    = "Hands, Tom O and Helled, R",
  journal   = "Mon. Not. R. Astron. Soc.",
  publisher = "Oxford Academic",
  volume    =  509,
  number    =  1,
  pages     = "894--902",
  month     =  oct,
  year      =  2021,
  language  = "en"
}

@ARTICLE{Turrini2021,
  title    = "Tracing the Formation History of Giant Planets in Protoplanetary
              Disks with Carbon, Oxygen, Nitrogen, and Sulfur",
  author   = "Turrini, D and Schisano, E and Fonte, S and Molinari, S and
              Politi, R and Fedele, D and Panić, O and Kama, M and Changeat, Q
              and Tinetti, G",
  journal  = "Astrophys. J.",
  volume   =  909,
  number   =  1,
  pages    =  40,
  year     =  2021
}

@ARTICLE{Shibata2022b,
  title     = "The origin of the high metallicity of close-in giant exoplanets -
               {II}. The nature of the sweet spot for accretion",
  author    = "Shibata, S and Helled, R and Ikoma, M",
  journal   = "Astron. Astrophys. Suppl. Ser.",
  publisher = "EDP Sciences",
  volume    =  659,
  pages     = "A28",
  month     =  mar,
  year      =  2022,
  language  = "en"
}

@ARTICLE{Knierim2022,
  title     = "Constraining the origin of giant exoplanets via elemental
               abundance measurements",
  author    = "Knierim, H and Shibata, S and Helled, R",
  journal   = "Astron. Astrophys. Suppl. Ser.",
  publisher = "EDP Sciences",
  volume    =  665,
  pages     = "L5",
  month     =  sep,
  year      =  2022,
  language  = "en"
}

@ARTICLE{Pacetti2022,
  title     = "Chemical Diversity in Protoplanetary Disks and Its Impact on the
               Formation History of Giant Planets",
  author    = "Pacetti, Elenia and Turrini, Diego and Schisano, Eugenio and
               Molinari, Sergio and Fonte, Sergio and Politi, Romolo and
               Hennebelle, Patrick and Klessen, Ralf and Testi, Leonardo and
               Lebreuilly, Ugo",
  journal   = "ApJ",
  publisher = "IOP Publishing",
  volume    =  937,
  number    =  1,
  pages     =  36,
  month     =  sep,
  year      =  2022,
  language  = "en"
}

@ARTICLE{Seligman2022,
  title     = "Inferring Late-stage Enrichment of Exoplanet Atmospheres from
               Observed Interstellar Comets",
  author    = "Seligman, Darryl Z and Becker, Juliette and Adams, Fred C and
               Feinstein, Adina D and Rogers, Leslie A",
  journal   = "ApJL",
  publisher = "IOP Publishing",
  volume    =  933,
  number    =  1,
  pages     = "L7",
  month     =  jun,
  year      =  2022,
  language  = "en"
}

@ARTICLE{Rodet2023,
  title     = "Planet-driven scatterings of planetesimals into a star:
               probability, time-scale, and applications",
  author    = "Rodet, Laetitia and Lai, Dong",
  journal   = "Mon. Not. R. Astron. Soc.",
  publisher = "Oxford Academic",
  volume    =  527,
  number    =  4,
  pages     = "11664--11684",
  month     =  dec,
  year      =  2023,
  language  = "en"
}

@ARTICLE{Shibata2024,
  title     = "Fate of a remnant solid disk around an eccentric giant planet",
  author    = "Shibata, S and Helled, R",
  journal   = "Astron. Astrophys.",
  publisher = "EDP Sciences",
  volume    =  689,
  pages     = "A26",
  month     =  sep,
  year      =  2024
}

@INPROCEEDINGS{Manara2023,
       author = {{Manara}, C.~F. and {Ansdell}, M. and {Rosotti}, G.~P. and {Hughes}, A.~M. and {Armitage}, P.~J. and {Lodato}, G. and {Williams}, J.~P.},
        title = "{Demographics of Young Stars and their Protoplanetary Disks: Lessons Learned on Disk Evolution and its Connection to Planet Formation}",
    booktitle = {Protostars and Planets VII},
         year = 2023,
       editor = {{Inutsuka}, S. and {Aikawa}, Y. and {Muto}, T. and {Tomida}, K. and {Tamura}, M.},
       series = {Astronomical Society of the Pacific Conference Series},
       volume = {534},
        month = jul,
        pages = {539},
          doi = {10.48550/arXiv.2203.09930},
archivePrefix = {arXiv},
       eprint = {2203.09930},
 primaryClass = {astro-ph.SR},
       adsurl = {https://ui.adsabs.harvard.edu/abs/2023ASPC..534..539M}
}

@ARTICLE{Emsenhuber2021,
  title     = "The New Generation Planetary Population Synthesis ({NGPPS})-{I}.
               Bern global model of planet formation and evolution, model tests,
               and emerging planetary …",
  author    = "Emsenhuber, A and Mordasini, C and Burn, R and Alibert, Y and
               {others}",
  journal   = "Astronomy",
  publisher = "aanda.org",
  year      =  2021
}

@ARTICLE{Booth2017,
  title    = "Chemical enrichment of giant planets and discs due to pebble drift",
  author   = "Booth, Richard A and Clarke, Cathie J and Madhusudhan, Nikku and
              Ilee, John D",
  journal  = "Mon. Not. R. Astron. Soc.",
  volume   =  469,
  number   =  4,
  pages    = "3994--4011",
  year     =  2017
}

@ARTICLE{Booth2019,
  title    = "Planet-forming material in a protoplanetary disc: The interplay
              between chemical evolution and pebble drift",
  author   = "Booth, Richard A and Ilee, J D",
  journal  = "Mon. Not. R. Astron. Soc.",
  volume   =  487,
  number   =  3,
  pages    = "3998--4011",
  year     =  2019
}

@ARTICLE{Schneider2021a,
  title     = "How drifting and evaporating pebbles shape giant planets - {I}.
               Heavy element content and atmospheric {C}/{O}",
  author    = "Schneider, Aaron David and Bitsch, Bertram",
  journal   = "Astron. Astrophys. Suppl. Ser.",
  publisher = "EDP Sciences",
  volume    =  654,
  pages     = "A71",
  month     =  oct,
  year      =  2021,
  language  = "en"
}

@ARTICLE{Schneider2021b,
  title     = "How drifting and evaporating pebbles shape giant planets - {II}.
               Volatiles and refractories in atmospheres",
  author    = "Schneider, Aaron David and Bitsch, Bertram",
  journal   = "Astron. Astrophys. Suppl. Ser.",
  publisher = "EDP Sciences",
  volume    =  654,
  pages     = "A72",
  month     =  oct,
  year      =  2021,
  language  = "en"
}

@ARTICLE{Savvidou2023,
       author = {{Savvidou}, Sofia and {Bitsch}, Bertram},
        title = "{How to make giant planets via pebble accretion}",
      journal = {\aap},
         year = 2023,
        month = nov,
       volume = {679},
          eid = {A42},
        pages = {A42},
          doi = {10.1051/0004-6361/202245793},
archivePrefix = {arXiv},
       eprint = {2309.03807},
 primaryClass = {astro-ph.EP},
       adsurl = {https://ui.adsabs.harvard.edu/abs/2023A&A...679A..42S}
}

@ARTICLE{Alcala2017,
  title     = "{X}-shooter spectroscopy of young stellar objects in Lupus:
               Accretion properties of class {II} and transitional objects⋆",
  author    = "Alcalá, J M and Manara, C F and Natta, A and Frasca, A and Testi,
               L and Nisini, B and Stelzer, B and Williams, J P and Antoniucci,
               S and Biazzo, K and Covino, E and Esposito, M and Getman, F and
               Rigliaco, E",
  journal   = "Astron. Astrophys.",
  publisher = "EDP Sciences",
  volume    =  600,
  pages     = "A20",
  month     =  apr,
  year      =  2017
}
\bibliographystyle{aasjournalv7}



\end{document}